\documentclass[twocolumn,trackchanges]{aastex631}

\usepackage{CJK}
\usepackage{graphicx}
\usepackage{amsmath}
\usepackage{multirow}
\usepackage{enumitem}
\setlist{nosep}
\graphicspath{{graphics/}}

\newcommand{\petit}{\texttt{petitRADTRANS}}

\newcommand{\kms}{$\rm km\ s^{-1}$}
\newcommand{\kp}{$K_p$}
\newcommand{\dv}{$\Delta v_\text{pl}$}
\newcommand{\vsys}{$v_{\rm sys}$}
\newcommand{\kpvsys}{$K_p-v_{\rm sys}$}
\newcommand{\dphi}{$\Delta \phi$}

\newcommand{\ucla}{Department of Physics \& Astronomy, 430 Portola Plaza, University of California, Los Angeles, CA 90095, USA}

\shorttitle{Hot Jupiter detectability with HRS}
\shortauthors{Hong et al.}

\graphicspath{{./}{}}

\begin{document}
\begin{CJK*}{UTF8}{gbsn}

\title{Velocity shift and SNR limits for high-resolution spectroscopy of hot Jupiters using Keck/KPIC}

\correspondingauthor{Kevin S. Hong}
\email{kevinh29@ucla.edu}

% Hong
\author[0000-0002-1024-7627]{Kevin S. Hong}
\affiliation{\ucla}

% Finnerty
\author[0000-0002-1392-0768]{Luke Finnerty}
\affiliation{\ucla}

% Fitzgerald
\author[0000-0002-0176-8973]{Michael P. Fitzgerald}
\affiliation{\ucla}

\begin{abstract}
\noindent
High-resolution cross-correlation spectroscopy (HRCCS) is a technique for detecting the atmospheres of close-in planets using the change in the projected planet velocity over a few hours. 
To date, this technique has most often been applied to hot Jupiters, which show a large change in velocity on short timescales.
Applying this technique to planets with longer orbital periods requires an improved understanding of how the size of the velocity shift and the observational signal-to-noise ratio impact detectability.
We present grids of simulated Keck/KPIC observations of hot Jupiter systems, varying the observed planet velocity shift and signal-to-noise ratio (SNR), to estimate the minimum thresholds for a successful detection.
{These simulations realistically model the cross-correlation process, which includes a time-varying telluric spectrum in the simulated data and data detrending via PCA.}
We test three different planet models based on an ultra-hot Jupiter, a classical hot Jupiter, and a metal-rich hot Saturn.
For a ${6}\sigma$ detection suitable for retrieval analysis, we estimate a minimum velocity shift of $\Delta v_\text{pl} \sim {30, 50, 60}$ \kms, compared to an instrumental resolution of 9 \kms, {and minimum SNR $\sim 370, 800, 1200$} for the {respective planet models.} % ultra-hot Jupiter model and $\Delta v_\text{pl} \sim 40$ \kms\ the other two planet models. 
% The minimum SNR depends strongly on the planet model and the star-to-planet photometric contrast ratios.
We find that reported KPIC detections to-date fall above {or near} the ${6}\sigma$ limit. 
These simulations can be efficiently re-run for other planet models and observational parameters, which can be useful in observation planning and detection validation.

\end{abstract}


% OUTLINE
% > Introduction
%  - explain how high-res emission spec complements low/med-res spec, what info can we get, compare with transimission spec
%  - Keck/KPIC: progress made, 
%  - limitations on detection strength
%     - what to expect for min kp/dv
%  - next sections...

% > Simulations
%  - explain basis of simulations
%     - what systems? what are their properties?
%         - TODO: separate physical/simulation table, also add more props including star
%     - maybe what to expect for each one based on those properties?
%     - what kind of grid: KP-SNR, DV-SNR
%  - planet models: go over how planet models are made for simulations
%  - stellar and telluric models: go over how stellar and telluric models are made
%  - simulated observations: how models are combined to get simulations
%  - HRCCS: how simulated observations are cross-correlated with (same) models

% > Results
%  - 6 main detection grid plots (kp-snr / dv-snr for each system)
%  - maybe example sub-grid of kp-vsys plots for varying kp-snr (big steps)
%  - compare with past results for WASP33 / HD189
%  - 
% > Discussion
% talk about tellurics, other effects we ignored
% parameterization, how it changes with kp/vp/snr/dphi/contrast
% > Conclusion

\section{Introduction} \label{sec:intro}
High-resolution cross-correlation spectroscopy (HRCCS) using ground-based telescopes has proven to be an effective technique in characterizing exoplanet atmospheres \citep[e.g.][]{line2021, pelletier2021, finnerty_keckkpic_2023, ramkumar_high-resolution_2023, finnerty_atmospheric_2024}.
HRCCS serves as an important complement to low- to medium-resolution transmission spectroscopy using space-based telescopes \citep{lustig-yaeger_jwst_2023, august_confirmation_2023}, allowing for the observation of non-transiting planets \citep[e.g.][]{finnerty2025a} and direct measurements of atmospheric winds and circulation patterns.
By resolving large numbers of spectral lines, HRCCS is capable of unambiguous molecular identification, measurement of atmospheric abundance ratios, and constraining vertical temperature structure \citep{brogi_high-resolution_2021, birkby_exoplanet_2018}.

HRCCS leverages the orbital motion of a close-in planet over several hours to isolate Doppler-shifted lines in the planet's spectrum from the relatively static stellar and telluric lines.
The residual spectrum after removing static features can then be compared with model planet spectra, which can be used to calculate the log-likelihood for Bayesian retrievals \citep[e.g.][]{brogi_retrieving_2019, gibson_detection_2020}.
This technique is effective for hot Jupiters that orbit close to their host star with a short orbital period for which we can observe a sufficient velocity shift in a single night.

{Data detrending processes used to remove time-varying telluric features necessitate a significant change in the projected planetary velocity.}
% This requirement of a significant change in the projected planetary velocity is compounded by the data detrending processes used to remove telluric features. 
These techniques, such as PCA and SysRem \citep{tamuz2005}, assume the observed spectrum is static except for planet features, and can significantly distort the original planet spectrum if the planet velocity shift is small relative to the instrumental resolution. 
This can decrease the detection significance or preclude detection entirely for small velocity shifts. 
Thus, a reliable detection of an exoplanet with HRCCS requires both sufficiently low noise and a sufficiently large planet velocity shift over the observation such that the planet signal can be accurately isolated and measured. 

For high-resolution transmission spectroscopy, the existence of planet-free out-of-transit frames can provide a sufficient basis for detrending to detect even sub-pixel velocity shifts of the planet during transit \citep{cheverall_feasibility_2024}. 
However, this is not applicable for emission spectroscopy of non-transiting planets. 
In this case, we must rely only on the planet velocity shift and the signal to noise ratio. 
The planet velocity $v_\text{pl}$ at any particular orbital phase $\phi$ is given by:
    \begin{equation} \label{eq:vpl_frame}
        v_\text{pl} = K_p \sin(2 \pi \phi) + v_\text{sys} - v_\text{bary}
    \end{equation}
where $K_p$ is the planet radial velocity semi-amplitude, $v_\text{sys}$ is the radial velocity of the planet system, and $v_\text{bary}$ is the barycentric velocity of Earth, {which is subtracted from the radial velocity}. 
Assuming constant systemic and barycentric velocities, the total change in planet velocity over an observed phase range $\Delta \phi = \phi_2 - \phi_1$ is reduced to 
    \begin{equation} \label{eq:vpl}
        \Delta v_\text{pl} = K_p [\sin(2 \pi \phi_2) - \sin(2 \pi \phi_1)]
    \end{equation}
Thus, the total planet velocity shift depends on the planet radial velocity semi-amplitude \kp\ and the total observed phase range \dphi{}. 
{Setting the change in $v_\text{bary}$ to 0 is not expected to significantly impact subsequent analysis. In real observations, the change in $v_\text{bary}$ over a single night is a few hundred $\rm m\ s^{-1}$ at most, compared to the 3--9 \kms\ spectral resolution HRCCS instruments, and is incorporated into the forward model of the system. }

Following \cite{birkby_exoplanet_2018} and \cite{snellen_combining_2015}, we can write to first-order the total signal-to-noise ratio (SNR) of the planet spectrum:
    \begin{equation} \label{eq:snr}
        \rm SNR_{planet} = \left(\frac{S_p}{S_\star}\right) SNR_{star} \sqrt{N_{lines}}
    \end{equation}
where $\rm {S_p}/{S_\star}$ is the planet-to-star flux ratio, $\rm SNR_{star}$ is the total signal-to-noise ratio of the host star which goes to $\sqrt{S_\star}$ in the bright/photon-limited regime, and $\rm N_{lines}$ is the number of resolved lines detected in the observed wavelength range.
We have neglected noise from the sky and telescope background, read-out noise, and the dark current from the detector. 
This effectively assumes our observations are in the photon-limited regime where the noise budget is dominated by the high flux of the star, which is generally appropriate for HRCCS targets. 

In this paper, we estimate the minimum planet velocity shift and SNR required to detect hot Jupiters with the HRCCS technique using simulated high-resolution planet spectra based on Keck/KPIC. 
In Section \ref{sec:sims}, we detail the method for simulating exoplanet observations for three different planet models and determining detection strengths. 
We present the simulation results in Section \ref{sec:results} and discuss these findings in Section \ref{sec:disc}. 
Finally, we summarize our work in Section \ref{sec:conc}.

\section{Simulations}\label{sec:sims}
We consider three planet models: an ultra-hot Jupiter system based on WASP-33 b \citep{finnerty_keckkpic_2023}, a hot Jupiter system based on HD 189733 b \citep{finnerty_atmospheric_2024}, and a metal-rich hot Saturn system based on HD 149026 b \citep{bean_high_2023}. 
The properties of these systems and the planet models are provided in Table \ref{tab:plmodels}.
% We discuss the planet models further in the next section.

\begin{deluxetable*}{ccccc}
    \tablehead{\colhead{Name} & \colhead{Symbol} & WASP-33 b & HD 149026 b & HD 189733 b}
    \startdata
        % $T_1$ [K] & & 1425 & 850 \\
        % $T_2$ [K] &  & 1500 & 900 \\
        % $T_3$ [K] &  & 1600 & 950 \\
        % $T_4$ [K] &  & 2000 & 1000 \\ 
        % $T_5$ [K] &  & 2100 & 1050 \\
        % $T_6$ [K] &  & 2150 & 1300 \\
        % $T_7$ [K] & - & - & 1500 \\
        % $T_8$ [K] & - & - & 1600 \\
        % $T_9$ [K] & - & - & 1650 \\
        Period [days] & P & 1.21983821$^a$ &  2.8758831$^a$ & 2.21857480$^a$ \\
        Semi-major axis [AU]  & $a$ & 0.0259$^b$  & 0.0432$^c$ & 0.031$^d$\\
        Inclination [deg] & $i$     & 87.7$^b$  & 84.55$^b$  & 85.71$^b$\\
        Planet RV semi-amplitude [\kms]  & $K_p$    & 230.9 & 54.6 & 153 \\
        % $v_\textb$ [\kms]    & -0.3 \\
        Stellar Radius [$R_\odot$] & $\rm R_*$ & 1.55$^b$ & 1.41$^b$ & 0.75$^b$ \\
        Planet Radius [$R_\textrm{J}$] & $\rm R_p$ & 1.6$^b$ & 0.74$^b$ & 1.13$^b$ \\
        Planet Mass [$M_\textrm{J}$] & $\rm M_p$ & 1.17$^b$ & 0.38$^b$ & 1.13$^b$ \\
        \hline
        log infrared opacity [$\rm cm^{2} g^{-1}$] & $\log \kappa_{\rm IR}$ [$\rm cm^{2} g^{-1}$] & -1.4 & -1.0 & -1.0 \\
        log infrared/optical opacity & $\log \gamma$ & -0.55 & -1.4 & 0.5 \\
        Equilibrium temperature [K] & $\rm T_{equ}$  & 2700 & 1640 & 800 \\
        log cloud-top pressure [bar] & $\log \rm P_{cloud}$  & - & - & -1.0 \\
        Rotational velocity [\kms] & $v_\text{rot}$ & 4  & 4 & 4 \\
        log H$_2$O mass-mixing ratio & log H$_2$O & -4.1 & -2.3 & 1.7 \\
        log CO mass-mixing ratio & log CO & -1.1 & -1.0 & -1.8 \\
        % $\rm \log CH_4\ MMR$ & - & - &  \\
        % $\rm \log NH_3\ MMR$ & - & - & -5  \\
        % $\rm \log HCN\ MMR$ & - & - & -5 \\
        log OH mass-mixing ratio & log OH & -2.1 & - & - \\
        % $\rm \log SiO\ MMR$ &  & -1.5 & - \\
        log H$_2$S mass-mixing ratio & log H$_2$S & - & -3 & - \\
        % $\rm \log C_2H_2\ MMR$ & - & - & -6 \\
        % $\rm \log CO_2\ MMR$ &  & -3.4 & -  \\
        log CO isotopologue ratio & $\rm \log \frac{^{13}CO}{^{12}CO}$ & -1.7 & -1.6 & -0.8 \\
        log H$_2$ mass-mixing ratio & $\log \rm H_2$ & -0.14 & -0.4 & -0.36 \\
        Scale factor & scale & 1.0 & 1.0 & 4.9
    \enddata 
    \caption{List of parameter values for the planet model of each system. $^a$ \citet{exofop2022}, $^b$ \citet{stassun_accurate_2017}, $^c$\citet{ment_radial_2018}, $^d$ \citet{rosenthal_california_2021}}
    \label{tab:plmodels}
\end{deluxetable*}

\subsection{Keck/KPIC}

Our simulations are based on previous HRCCS results using the Keck Planet Imager and Characterizer \cite[KPIC;][]{nirspec, nirspecupgrade, nirspecupgrade2, kpic, echeverri_phase_2022, jovanovic2025} in the $K$-band. KPIC consists of a series of upgrades to the Keck II adaptive optics system and NIRSPEC high-resolution spectrograph to enable diffraction-limited high-resolution spectroscopy in the $K$- and $L$-bands. Originally developed to enable spectroscopy of directly-imaged substellar companions, the improved wavelength and blaze function stability offered by the fiber feed also offers significant advantages for HRCCS applications. To-date, KPIC observations have led to published detections of four hot Jupiters \citep{finnerty_keckkpic_2023, finnerty_atmospheric_2024, finnerty2025a, finnerty2025b}, and analyses of several more targets are in preparation. 

KPIC $K$-band observations cover $\sim1.9-2.5$ $\mu$m in 9 spectral orders. The three bluest orders are heavily contaminated by telluric CO$_2$ absorption and are not useful for analysis. In this work, we consider the four reddest orders of the $K$-band, covering $\sim2.2-2.5$ $\mu$m, with significant gaps (see Figure \ref{fig:spec} for the wavelength coverage) at a spectral resolution $R = \lambda/\Delta\lambda\sim35,000$. These orders cover the 2.3 $\mu$m CO bandhead, and also have significant H$_2$O opacity \citep{finnerty_keckkpic_2023, finnerty_atmospheric_2024}. Two additional orders can be calibrated to sufficient precision for HRCCS applications, but generally do not contribute significantly to HRCCS detections due to low molecular opacities in these bands \citep{finnerty2025a, finnerty2025b}, and are therefore omitted from our simulations. 

Compared to other instruments used for HRCCS, KPIC has a relatively low spectral resolution ($\sim9$ \kms). The HRCCS technique relies on the projected planetary velocity changing by multiple times the instrument resolution, and therefore we expect that the required minimum velocity shift will scale proportionally to the instrument resolution. While KPIC has a relatively low throughput from the top of the atmosphere \citep[$\sim2-3\%$][]{echeverri_phase_2022}, the 10m aperture of the Keck telescope offers a sensitivity comparable to higher-throughput instruments on smaller telescopes (such as Gemini/IGRINS). The simulations described below vary the total signal-to-noise ratio (SNR), which is instrument-agnostic. More sensitive instruments will reach a given SNR faster, but the detectability for a fixed spectral resolution is expected to be a function of the SNR, independent of the instrument.

% we simulate a grid of observed spectra. Given observation data, stellar model, and planet model for each system, a simulated observed spectrum can be created by combining 

% - how were simulations done

% - Figure comparing Finnerty 2023 results for WASP-33 with this simulation code, should be similar

% - kpvsys : kp(-500,500), vsys(-100,100)

% - KP0(0, 250km/s)

% - deltav(10,150) based on obs time

% - snr(10,200)

% - 3 systems: UHJ (WASP33b), HJ (HD189733b, like WASP77), HN , metal-rich from JWST? (use HD189733 with high Z, like HD149026)

% - get updated planet model

% - no fringing (too complicated with , yes Gaussian noise

\subsection{Planet Models}
All planet models were generated using \petit\ \citep{prt:2019,prt:2020} with constant vertical mixing profiles and the pressure-temperature profile parameterization from \citet{guillot_radiative_2010} with $\rm T_{int} = 100\ \textrm{K}$. 
Planet model parameters are listed in Table \ref{tab:plmodels}.
The radiative transfer calculation was performed with 80 pressure layers ranging from $10^2$ to $10^{-6}$ bar and did not account for scattering.
To simulate an ultra-hot Jupiter, we chose parameters based on the WASP-33 b results from \citet{finnerty_keckkpic_2023}.
For the case of a high-metallicity atmosphere, we use parameters based on those reported in \citet{bean_high_2023} for HD 149026 b. 
Finally, to simulate a cooler hot Jupiter, we use the maximum-likelihood parameters for HD 189733 b reported in \citet{finnerty_atmospheric_2024}

\begin{figure}
    \centering
    \includegraphics[width=\columnwidth]{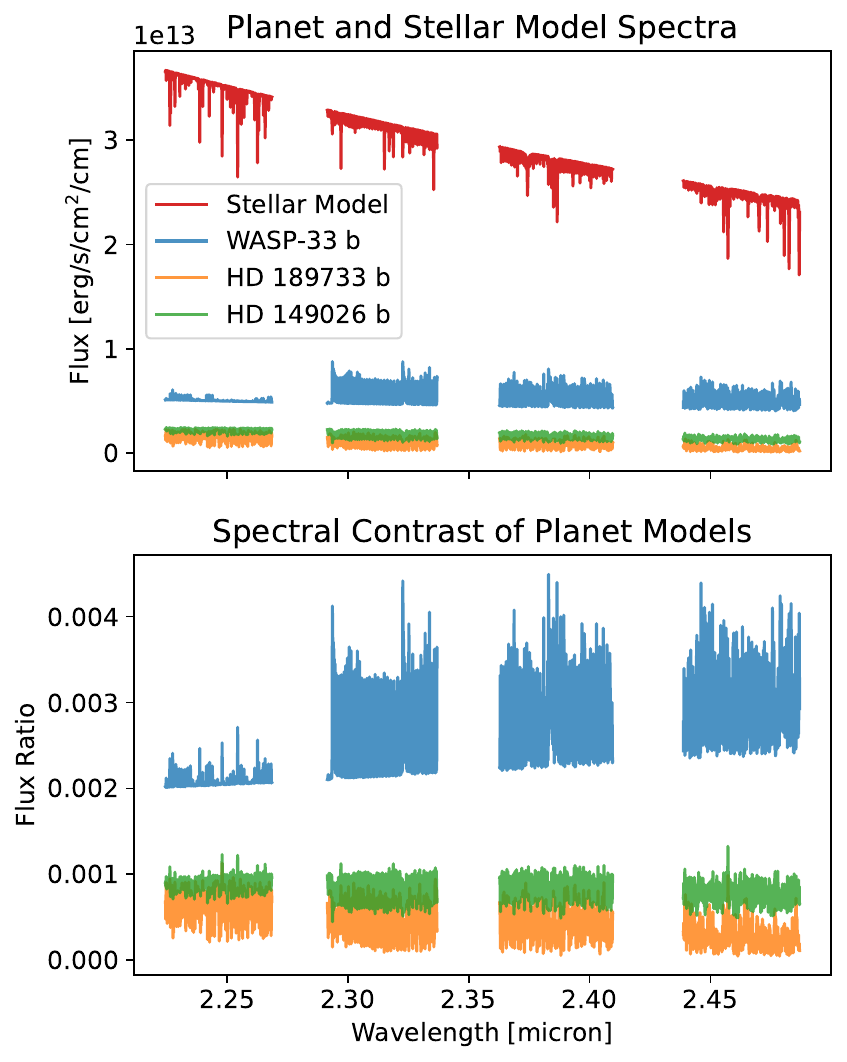}
    \caption{(Top) Planet and stellar model spectra in four KPIC orders used for simulating spectra. (Bottom) Planet-to-star spectroscopic flux ratio for each planet model. We assumed in our simulations that the planet and star radii were the same for all systems, resulting in a planet-to-star area ratio of 0.0145. The WASP-33 b model has the greatest flux ratio among all planet models and shows a strong thermal inversion and strong continuum. The HD 189733 b and HD 149026 b models have a similar continuum, but the former shows slightly stronger spectral features relative to the continuum.}
    \label{fig:spec}
\end{figure}

The tabulated opacities H$_2$O, OH, $^{12}$CO, and $^{13}$CO were previously described in \citet{finnerty2023}. 
For H$_2$O, we used the \cite{polyansky2018} partition function and HITEMP 2010 linelist \citep{hitemp2010}. 
For both CO isotopologues, we used the HITEMP 2019 lists \citep{hitemp2020} and the \citet{li2015} partition function. 
For OH, we used the partition function from \citet{YOUSEFI2018} and HITEMP 2020 linelist \citep{hitemp2020}. 
For H$_2$S, we used the HITRAN 2012 linelist \citep{rothman2013}. 
Opacities were generated using ExoCross \citep{exocross2018} following instructions in the \petit{} documentation.
Collisionally induced absorption (CIA) opacity from $\rm H_2$ and He are included, but scattering is only included for the HD 189733 b model, which includes a grey cloud deck as described in \citet{finnerty_atmospheric_2024}. 
The HD 189733 b model is also scaled by a factor of 4.9 based on the cooler-than-expected retrieved $P-T$ profile, {which results from limited sensitivity to absolute temperature in the free-retrieval framework} as discussed in \citet{finnerty_atmospheric_2024}. 
For the WASP-33 b model, half of the total hydrogen mass is taken to be in the form of HI rather than H$_2$. 
The planet model spectra are plotted in Figure \ref{fig:spec}.
Note that the WASP-33 b model shows a strong thermal inversion and strong continuum and that the HD 189733 b and HD 149026 b models have a similar continuum level.

% \subsection{Stellar and Telluric Models}
% For HD 189733 and HD 149026, we use a PHOENIX stellar model \citep{phoenix} with $\text{T}_{\text{eq}} = 4900\ \text{K}$, $\log g = 4.5$, and $\text{[Fe/H]} = 0.0$. For WASP-33, we use a PHOENIX stellar model with $\text{T}_{\text{eq}} = 7400\ \text{K}$, $\log g = 4.0$, and $\text{[Fe/H]} = 0.0$. We apply rotational broadening based on the expected $v \sin i$ for each system and limb darkening coefficients from \cite{sing2010}.

\subsection{Simulated Observations}
We use a PHOENIX stellar model \citep{phoenix} with $\text{T}_{\text{eq}} = 5800\ \text{K}$, $\log g = 4.5$, and $\text{[Fe/H]} = 0.0$ for each system. 
We apply rotational broadening with $v \sin i = 2$ \kms\ and limb darkening coefficients from \cite{sing2010}. 
While in reality the simulated planets orbit very different host stars, we chose to use a single stellar model based on a G-type star in order to isolate the impact of differences in the planetary spectrum on detectability.
For both planet and stellar models, we follow past observations using KPIC and use only the last four NIRSPEC orders ${\sim}2.2-2.5\ \rm{\mu m}$  due to telluric contamination and inaccurate wavelength calibration in the first few orders \citep{finnerty_keckkpic_2023, finnerty_atmospheric_2024}.

Our simulations vary the total planet velocity shift \dv\ and the total SNR of the time series. 
For all simulations, we generate a simulated time series of spectra with evenly spaced orbital phases of step size $\delta\phi = 0.00174$ ($0.0109$ radians).
This is based on an assumed orbital period of 2 days for all systems and a constant integration time of 300 seconds per frame.
For $K_p = 150$ \kms{}, the average planet velocity shift within a frame is ${\sim}1.5$ \kms{}.
This shift is substantially smaller than the $\sim$9 \kms\ instrument resolution, and we do not simulate smearing of the planetary spectrum due to motion within a single frame.

Based on Equation \ref{eq:vpl}, the total velocity shift of the planet over an observation depends on both the planet radial velocity semi-amplitude \kp\ and the total observed phase range \dphi.
Thus, we consider two methods in varying the planet velocity shift for the simulations:
\begin{itemize}
    \item In the ``fixed-phase'' grid case, we use an evenly-spaced orbital phase range $\phi = 0.52-0.58$ ($\Delta \phi = 0.06$) for a total of 35 frames for all simulations. 
Then, we produce the given simulated planet velocity shift \dv\ by varying \kp\ using Equation \ref{eq:vpl}. 
    \item In the ``fixed-\kp{}'' grid case, we set a consistent $K_p=150$ \kms\ for all simulations. 
Then, we generate a variable-size array of observed phases that begins at $\phi_1 = 0.52$ with the step size above, where final phase and total number of frames is determined by the given velocity shift \dv{}.
\end{itemize}

The total SNR for a given simulation is divided between each frame, given by:
\begin{equation}
    \rm SNR_{total} = SNR_{frame} \sqrt{N_{frames}}
\end{equation}
where $\rm N_{frames}$ is the number of frames. 
Hence, for simulated spectra with more frames (in the fixed-\kp{} grid case), the per-frame SNR is lower, but the total SNR remains constant. 

For both grid cases, \dv\ is varied from 2 \kms\ to 101 \kms\ in steps of 1 \kms{}.
Total SNR is varied from 15 to 1500 in steps of 15.
This results in a $100\times100$ grid of simulations for a given planet model and grid case.
The two grid cases are summarized in Table \ref{tab:grids}. 

\begin{deluxetable*}{ccccc}
    \label{tab:grids}
    \caption{The simulation parameters for the two grid cases. The SNR and \dv\ ranges are the same for both grid cases. We use a step size of 1 \kms\ for \dv\ and a step size of 15 for SNR. In the fixed-phase grid case, we use a consistent orbital phase range from $\phi = 0.52$ to $\phi = 0.58$ (total $\Delta \phi = 0.06$) while varying \kp\ to produce the simulated \dv. In the fixed-\kp{} case, we keep a constant $K_p = 150$ \kms\ and vary the total phase change \dphi\ to produce the total velocity shift (starting at $\phi_1 = 0.52$).}
    \tablehead{\colhead{Grid Case} & SNR & \dv\ [\kms] & \kp\ [\kms] & \dphi}
    \startdata
        Fixed-phase & [15, 1500, 15] & [2, 101, 1] & [6, 283, 2.8] & 0.06 ($0.52-0.58$)\\
        Fixed-\kp{} & [15, 1500, 15] & [2, 101, 1] & 150 &  [0.002, 0.127, 0.001]
    \enddata
\end{deluxetable*}

To simulate an observed spectrum, we first scale the planet model by the planet-to-star area ratio, using a consistent planet radius of 1.2 $R_{\rm J}$ and stellar radius of 1.0 $R_\odot$ for an overall geometric contrast factor of 0.0152. 
We assume a systemic velocity and barycentric velocity of zero for all frames so the stellar model spectrum has no change in Doppler shift over the observed time series. 
{The change in barycentric velocity over a single night is small compared to the 3--9 \kms{} spectral resolution HRCCS instruments so setting the barycentric velocity to 0 will not significantly impact the analysis.}
The planet model is Doppler shifted based on the planet velocities given by Equation \ref{eq:vpl_frame} for each frame and interpolated onto the observed wavelength grid \citep{finnerty_keckkpic_2023}.
{After the stellar and planet models together, we multiply a time-varying telluric transmission model spectrum onto the combined spectrum.}
{The telluric spectrum is generated using PSG \citep{psg}, which is varied based on an airmass that ranges from 1 to ${\sim}1.4$ over the observed phase, representing a typical change over a night of observing.}
{By adding this simple time-varying telluric model, we can apply PCA to investigate the impact of typical detrending processes on detection strengths.}
% The stellar and planet models are added together and a telluric transmission model spectrum generated using PSG \citep{psg} is multiplied to the final combined spectrum.
The model spectrum is convolved with a 1.3 pixel Gaussian filter to simulate the instrumental broadening of KPIC \citep{finnerty_keckkpic_2023, finnerty_atmospheric_2024}. 
To simulate the total observed SNR, the combined spectrum is first mean-subtracted, and Gaussian noise is added frame-wise to the spectrum based on the given total SNR divided over each frame ($\sigma \sim 1/{\rm SNR_{frame}};\ \rm SNR_{frame} = SNR_{total} / \sqrt{N_{frames}}$).

The data detrending process for the simulated spectra follows \citep{finnerty_keckkpic_2023}.
{This includes a time-series median division to remove static stellar features and principal component analysis (PCA) to remove time-varying features e.g. airmass and time-varying fringing effects from the optics of KPIC.}
{While we do not explicitly include the fringing effects, we still use PCA to remove the time-varying telluric features that we simulated, which could be augmented to account for fringing.}
{However, PCA can distort the planet signal, which is minimized by removing the smallest number of principal components (PCs) required to correct for instrumental/telluric systematics \cite{Finnerty2022}.}
{For our analysis, we remove 4 PCs to be consistent with previous KPIC analyses \citep{Finnerty2022, finnerty_keckkpic_2023}, but we note that removing only 2 PCs produced similar detection grids.}
% a time-series median division to remove static stellar features and .
% However, because we do not include the time-varying fringing effects associated with KPIC \citep{Finnerty2022}, it is not necessary to perform a principal component analysis (PCA) to remove these components and isolate the planet signal, as was necessary in \citet{finnerty_keckkpic_2023}. 
{We further} assume that any distortions from PCA are included in the simulated noise of the spectra and that the residuals after PCA are Gaussian-distributed, which appears to be a good assumption for the data previously reported in \citet{finnerty_keckkpic_2023}.

\subsection{Cross-Correlation and Log-Likelihood}
{For each simulated spectrum, we perform cross-correlation with a planet template model, which uses the same planet (and stellar) model used to generate the given spectrum.}
{We cross-correlate over a grid of \kp{} and \vsys{} values to produce a \kpvsys{} diagram that can be used to quantify detection strengths.}
{For a given \kp{} and \vsys{} in this grid, we Doppler shift the planet and stellar models according to Equation \ref{eq:vpl_frame} using the simulated observed phases for the simulation and combine the spectra.}
{The generation of the template spectrum is very similar to the simulation process, except it instead multiplies by a constant telluric spectrum and does not add any additional noise.}
{We subsequently process the template by median dividing and performing PCA, dropping 4 PCs, similar to the detrending of the simulated data.}
{At this point, the simulated data and the template spectrum can be used to calculate the log-likelihood values \citep{brogi_retrieving_2019} over the entire \kpvsys{} grid.} 
{For the \kpvsys{} diagram, we vary \kp{} between -300 to 300 \kms{} and vary \vsys{} between -40 to 40 \kms{}, both with a 4 \kms{} step size.}
{By generating a smaller \kpvsys{} diagram with a lower resolution, we can simulate detections at a faster speed, which we discuss further in Section \ref{sec:disc-simeff}.}

{Due to the detrending process and the dependence on the forward-model variance in the \citet{brogi_retrieving_2019} log-likelihood function, a feature at $K_p=0$ is produced in the \kpvsys{} diagram from the self-division of planet lines at low velocities.}
{To minimize the impact of this feature on our detection strength estimates, especially for low \dv{} simulations, we additionally process simulated \kpvsys{} diagrams by dividing each \kp{} column by the average value of the respective negative \kp{} column.}
{Doing so removes the feature near $K_p=0$ and mitigates artificially large detection strength estimates for low velocity simulations.}

\begin{figure}[h!]
    \centering
    \includegraphics[width=\linewidth]{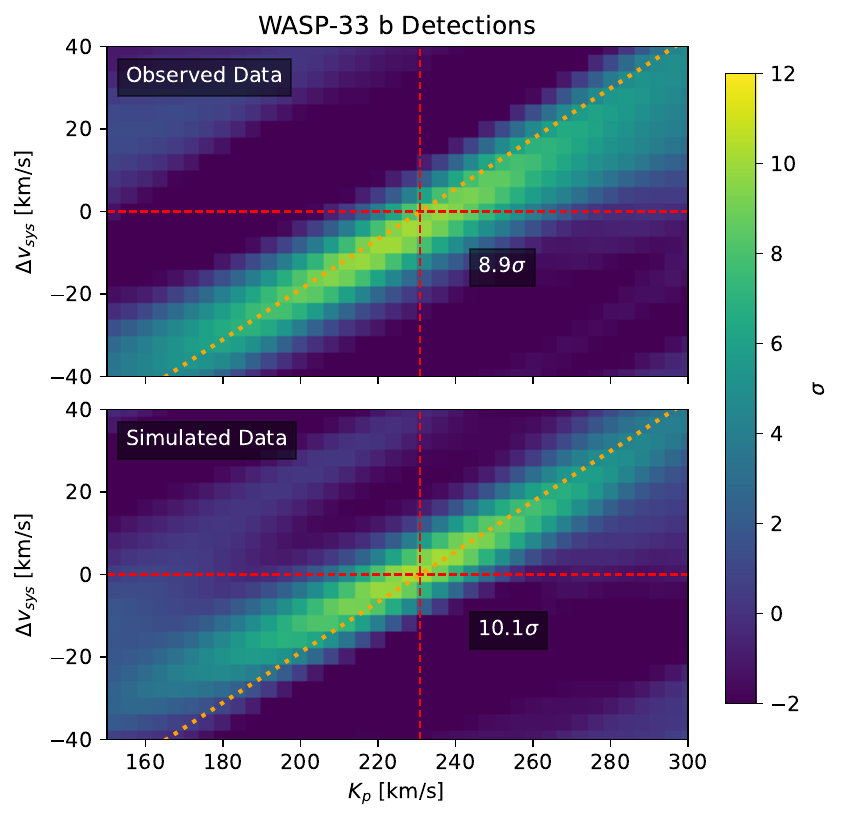}
    \caption{\kpvsys\ diagrams for the 2021 November 21 observations of WASP-33 b (top; \cite{finnerty_keckkpic_2023}) and for simulated WASP-33b data with observed parameters $\Delta v_\text{pl} = 140$ \kms and $\textrm{SNR} = 650$ that closely match observation (bottom). The red dashed lines mark the true velocities of WASP-33 b ($\Delta v_{\rm sys} = 0$, $K_p = 226$). The orange dashed line show the estimated degeneracy between $\Delta v_{\rm sys}$ and $K_p$ based on the observed and simulated phase range (Equation \ref{eq:vpl_frame}). Note that the colorbar scale is the same between both grids. The simulated data produces a detection strength estimate of ${10.1}\sigma$, which is {slightly larger} than the ${8.9}\sigma$ detection in the observed data.}
    \label{fig:wasp_kpvsys}
\end{figure}

In Figure \ref{fig:wasp_kpvsys}, we compare the \kpvsys{} diagrams  for real Keck/KPIC observations of WASP-33 b \citep{finnerty_keckkpic_2023} and for a simulated spectrum of WASP-33 b. 
The simulation uses the following parameters that closely match real observed values: $\Delta v_\text{pl} = 140$ \kms{}, $\textrm{SNR} = 650$, $K_p=226$ \kms{}, planet-to-star area ratio of ${\sim}0.0114$, phase range of 0.52--0.64, period of 1.22 days, and integration time of 250 seconds \citep{finnerty_keckkpic_2023}.

We can see clear detections in both \kpvsys\ diagrams in Figure \ref{fig:wasp_kpvsys}, represented by the peak in log-likelihood values at the true planet and systemic velocities (note that the same scale is used for both grids). 
We can also see the correlation between \kp\ and \vsys\ as expected based on Equation \ref{eq:vpl}. 
In Figure \ref{fig:wasp_kpvsys}, we calculate a detection strength of $8.9\sigma$ and $10.1\sigma$ for the observed and simulated data of WASP-33 b, respectively (see Section \ref{sec:ds}). 
The slightly higher detection strength in the simulated data suggests there may be some residual model mismatch in the observed case, even after atmospheric retrieval, which is not present in simulations.
By instead taking the maximum likelihood value near the peak to account for the model mismatch, we obtain a detection of $10.0\sigma$, which closely matches the simulated detection strength above.
Therefore, the simulated and observed detection strengths are in reasonably good agreement, indicating our simulation framework provides a reasonably accurate estimate of the detection strength, despite omitting instrumental systematics such as fringing.
% As expected, the simulation produces a {slightly larger} detection strength due to our simplifications in data detrending and perfect model matching.
Based on the simulations, a factor of ${\sim}1.1$ reduction in detection strength for real observations is plausible, but may depend on the exact difference in contrast between simulation and observation for a given planet.
The quantitative detection strength estimates from these \kpvsys\ diagrams are discussed in more detail in the next section. 
% The simulated data produces a detection strength estimate (see Section \ref{sec:ds}) that is ${\sim}{1.2}\sigma$ larger than that of the observed data.

\subsection{Detection Strength Estimation}\label{sec:ds}
\begin{figure}
    \centering
    \includegraphics[width=\linewidth]{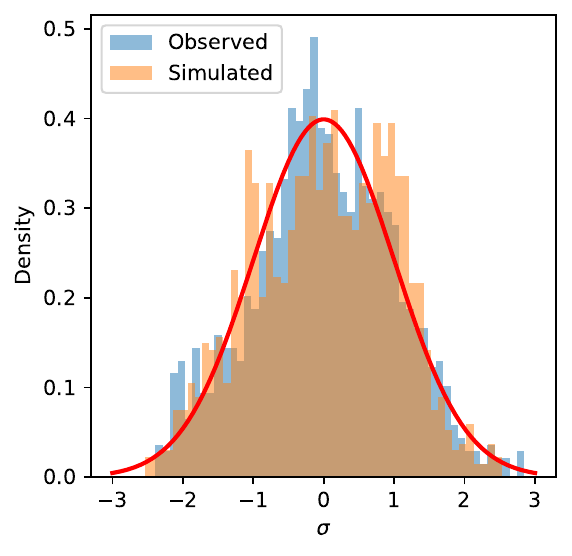}
    \caption{Histogram of off-peak log-likelihood values in the \kpvsys{} diagrams for the observed and simulated WASP-33 b data as in Figure \ref{fig:wasp_kpvsys}. The \kpvsys{} diagram is first shifted by the mean of off-peak values and scaled by the standard deviation of off-peak values as detailed in the text. A standard Gaussian is shown in red. We can see that the off-peak likelihood values are {approximately} Gaussian-distributed {for both the real and simulated data}.}
    \label{fig:noise}
\end{figure}

Quantifying the detection strength from a $K_p-v_\text{sys}$ diagram can be challenging due to the presence of non-Gaussian features far from the planet peak. 
These features can arise from telluric or stellar contamination or from the way the log-likelihood mapping is implemented \citep{finnerty_atmospheric_2024}, which makes defining a true ``noise" for signal-to-noise estimates difficult.  
To estimate the noise, we consider the log-likelihood values away from the peak value, or off-peak values, specifically in the non-physical velocity range $K_p < {-100}$ \kms{}. 
Given our simulation parameters, this region of \kpvsys{} space should not contain a detection and only include noise {while also avoiding the feature near $K_p=0$ from the detrending process}. 
% Because the planet radial velocity semi-amplitude \kp\ is defined to be positive, this region should not contain a detection and only include noise.
As seen in Figure \ref{fig:noise}, this region of \kpvsys\ space is dominated by Gaussian-distributed noise in the simulated spectrum \citep{brogi2012}. 
The mean of the off-peak values is subtracted from the entire \kpvsys{} diagram, which is then divided by the standard deviation of the off-peak region.
This scales the log-likelihood values to the noise far from the peak and represents the significance of the peak log-likelihood values compared to the {background} noise in the \kpvsys\ diagram, {which is quantified by $\sigma$ in this paper}. 

We consider two methods to quantify the detection strength. 
In the first method, we take the significance value at the true velocities (\kp\ and \vsys) of a given simulation.
In the second method, we take the maximum significance value in a diagonal area of the \kpvsys{} diagram that is close to the true velocity shift of the planet. 
Specifically, we take the maximum in the region that is within 10 \kms{} of the true velocity shift in every simulated frame, within 50 \kms{} of the true \kp{}, and within 50 \kms{} of the true \vsys{}.
In general, these two methods produce similar detection strength estimates for a sufficient \dv{} and SNR, where the maximum value is found very close to the true velocities, as in \kpvsys{} diagrams in Figure \ref{fig:wasp_kpvsys}.
However, at low \dv{} and low SNR, the maximum nearby detection strength may capture false detections that are the result of noise, artificially increasing the estimated detection strength. 
Therefore, for our main results, we use the detection strength at true velocity values because it produces more modest estimates compared to the maximum nearby detection strength.
We further discuss the differences between these two methods for determining the detection strength in Section \ref{sec:disc-maxsig}.

In many cases, especially in low-SNR data, there are often values of ${>}4\sigma$ throughout the \kpvsys\ space, which are false positives due to random noise in the data. 
We can see some of these false positive off-peak values in both observed and simulated \kpvsys\ diagrams in Figure \ref{fig:wasp_kpvsys}. 
Due to the presence of these spurious peaks away from the true velocity values, it is important to choose an appropriate lower bound at which a detection is significant.
In previous works, the planetary atmosphere is said to be detected if a high significance value, usually ${>}6\sigma$, is calculated at the true velocities \citep{brogi2019, finnerty_keckkpic_2023, finnerty_atmospheric_2024}.
However, the choice for a detection cutoff is arbitrary and depends on the given data and noise.
For our simulations, we adopt a detection cutoff of ${6}\sigma$ {because our simulations produced similar detections to real data given our realistic treatment of time-varying features and the detrending process.}
{However, our simulations still include optimistic assumptions about observed systems, such as assuming there is no model mismatch which will always impact real observations.}
% due to the optimistic assumptions in our simulations, which do not include the model mismatch which will always impact real observations.

\section{Results}\label{sec:results}
\subsection{$3\times3$ Detection Grid}
\begin{figure*}[ht!]
    \centering
    \includegraphics[width=\textwidth]{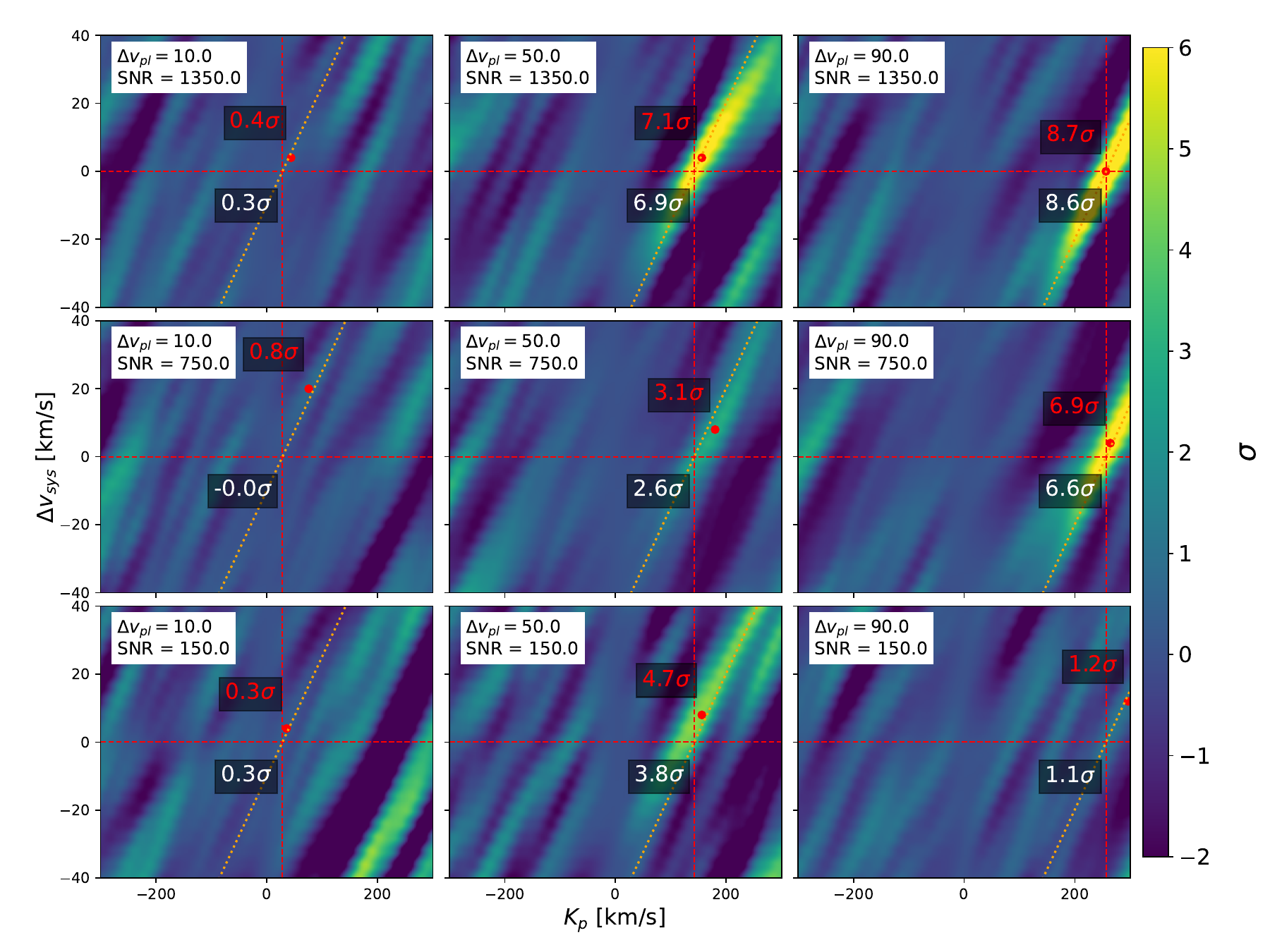}
    \caption{$3\times3$ grid of \kpvsys\ diagrams of simulated HD 189733 b detections for \dv\ = 10, 50, and 90 \kms\ and total SNR = 150, 750, and 1350. These diagrams are from the fixed-phase grid case using a consistent orbital phase range $\phi = 0.52 - 0.58$ $(\Delta \phi = 0.06)$. The color bar is scaled to a maximum of ${6}\sigma$ to emphasize the non-detections of the low \dv\ and SNR regime. The red dashed lines indicate the true \kp\ and \vsys\ for a given simulation, with the true velocity detection strength shown. The maximum nearby detection strength is shown as the red dot and red text. The orange dashed lines indicate the estimated correlation between \kp\ and \vsys\ given the \dv\ and phase range (Equation \ref{eq:vpl}). Note that the colorbar scale is the same for all diagrams. We can see that the two detection strength methods produce consistent results for high \dv{} and SNR, but deviates at lower \dv{} and SNR.}
    \label{fig:small_grid}
\end{figure*}

In Figure \ref{fig:small_grid}, we present a $3\times3$ grid of \kpvsys\ diagrams for the HD 189733 b simulations.
We show part of the fixed-phase simulation grid with \dv\ = 10, 50, and 90 \kms\ (\kp = 28.5, 142.4, and 256.4 \kms, respectively) and total SNR = 150, 750, and 1350. 
% The \kpvsys\ diagrams in Figure \ref{fig:small_grid} are scaled 
As expected, we see the strongest detection (${8.6}\sigma$) at \dv\ = 90 \kms\ and SNR = 1350 in the upper right \kpvsys\ diagram in Figure \ref{fig:small_grid}.
The detection strength decreases with decreasing \dv\ and SNR, until the planet becomes undetected ($<3\sigma$) in the \kpvsys\ diagrams at the lowest \dv\ and SNR, again consistent with expectations. 
The maximum nearby detection strength (red) is consistent with the true velocity detection strength for the significant detections in Figure \ref{fig:small_grid}, but overestimates detection strength in {some of} the non-detections. 

\subsection{$100\times100$ Detection Grids}\label{subsec:100x100}
\begin{figure*}[p!]
    \centering
    \includegraphics[width=0.8\textwidth]{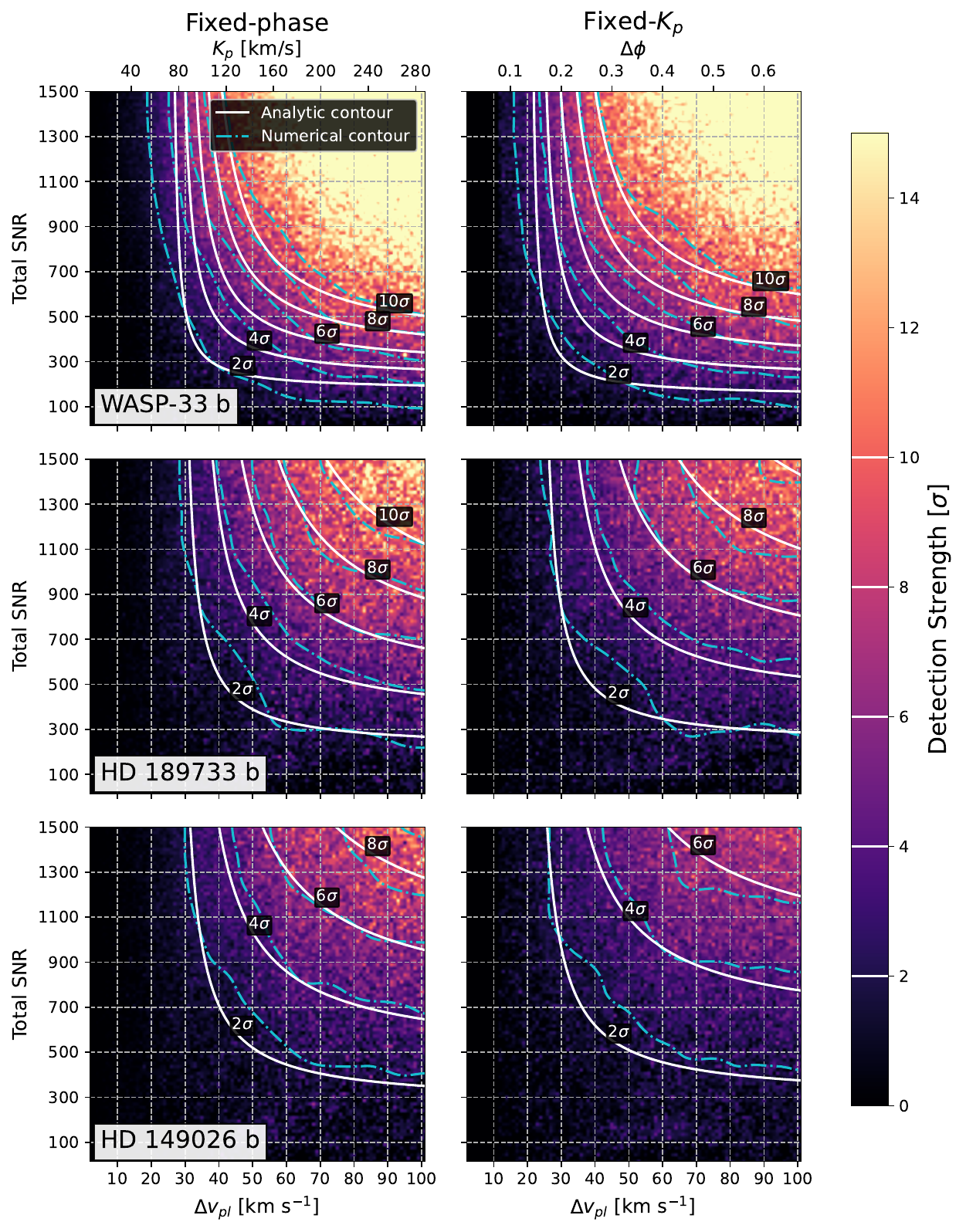}
    \caption{$100\times100$ detection strength grids for each planet model and grid case. The left column show the fixed-phase simulations and the right column show the fixed-\kp{} simulations. The first row shows the WASP-33 b grids, the middle row shows the HD 189733 b grids, and the last row shows the HD 149026 b grids. All grids have the same colorbar scale, with an upper bound of $15\sigma$ to highlight the contrast of lower values. We plot the best-fit contours based on Equation \ref{eq:contour_eq} (white) and the numerical contours from \texttt{matplotlib} (green), which correspond closely except in the intermediate region of \dv{} and SNR. The WASP-33 b model produces the strongest detection and thus lowest detection limits of all three planet models.}
    \label{fig:grids}
\end{figure*}

In Figure \ref{fig:grids}, we present the $100\times100$ grids of detection strength estimates for each planet model and grid case.
Each row presents a different planet model, and the left and right columns present the fixed-phase and fixed-\kp{} cases, respectively.
In each grid, a clear region of significant detection can be seen at high \dv\ and SNR.
The WASP-33 b model (ultra-hot Jupiter) in the top row of Figure \ref{fig:grids} demonstrates the largest region of ${>}{6}\sigma$ detections, while the HD 140926 b model (high-metallicity hot Saturn) is the least detectable of the three models considered.
Both the fixed-phase and fixed-\kp{} grids produce similar detection regions for the same planet model.

To estimate the lower limits of detectability, we fit the detection grids with a contour curve of the following functional form:
\begin{equation}\label{eq:contour_eq}
    D = \dfrac{(\Delta v_\text{pl} - A)(\textrm{SNR} - B)}{P \times \Delta v_\text{pl} + Q \times  \textrm{SNR}}
\end{equation}
where $D$ is the detection strength estimate, and $\{A, B, P, Q\}$ are parameters of the contour model.
In general, the first two parameters ($A, B$) determine the shift in contour asymptotes along \dv\ and SNR while the last two parameters ($P, Q$) determine the gradient of the contours, respectively.
This function is motivated by the apparent asymptotic nature of detection strengths at high \dv\ or high SNR for a given significance shown in Figure \ref{fig:grids}. 
We use the \texttt{curve\_fit} function from \texttt{scipy} to fit the contour function to grid data.

We compare this parameterization with the numerical contours calculated using \texttt{matplotlib} on the grids shown in Figure \ref{fig:grids} after smoothing with a 4-pixel Gaussian kernel.
The best-fit analytic contours (solid white) and numerical contours (dashed cyan) are overplotted on the grids in Figure \ref{fig:grids}.
We can see that the analytic contours are consistent with the numerical contours at the high velocity shift and high SNR asymptotic limits in each grid in Figure \ref{fig:grids}, with slight deviation at high-SNR asymptotes, where the analytic estimates are slightly more conservative than the numerical contours.
Therefore, we can use the asymptotic limits as approximate estimates for the detection strength of observations in the high-\dv{} or high-SNR regime.
However, the analytic contour deviates from the numerical contour in the intermediate, concave region in every grid, producing slightly lower limits.
The lower limits in this intermediate region by the analytic fit predicts that planets are easier to detect than the numerical contours suggest.
Despite these deviations, the similar behavior between the best-fit and numerical contours at the limits of \dv\ and SNR justifies the use of the contour function to estimate detection strength limits.

To estimate the ${6}\sigma$ detection limit on \dv{} in the high-SNR limit, we use Equation \ref{eq:contour_eq} at the maximum SNR of the grid ($\textrm{SNR} = 1500$).
This limit represents bright planets with a high spectral contrast but with a slow orbit or brief observation time.
By calculating at the maximum SNR of the grid rather than taking $\textrm{SNR} \to \infty$, we produce more a modest and realistic bound for the minimum velocity shift because the grid represents typical values for SNR and \dv{}.
Similarly, in the high-\dv{} limit, we use the maximum velocity shift of the grid ($\Delta v_\text{pl} = 100$ \kms) to calculate the ${6}\sigma$ detection limit on SNR in the limiting case of a dim but fast-orbiting planet.
We present the resulting ${6}\sigma$ detection limits for velocity shift \dv\ and SNR for each planet model and grid case in Table \ref{tab:mins}.
The detection limits are calculated using the fit parameter values from the contour function in Equation \ref{eq:contour_eq}.
We also compute errors in the asymptotic limits using the approximate covariance matrix of the fit parameter values.

\begin{table}[h]
    \centering
    \caption{Lower limits for the planet velocity shift \dv\ and SNR for a ${6}\sigma$ detection strength using the contour fit function in \ref{eq:contour_eq}. Errors are calculated using the approximate analytic covariance matrix on the fit parameters. }
    \label{tab:mins}
    \begin{tabular}{c|c|cc}
    %  & & \multicolumn{1}{c}{Minimum \dv}   & \multicolumn{1}{c}{Minimum SNR} \\ 
    % Planet Model & Grid case & \multicolumn{1}{c}{at $4\sigma$}  & \multicolumn{1}{c}{at $4\sigma$} \\ \hline
    % \multicolumn{1}{c|}{Planet Model} & \multicolumn{1}{c|}{Grid case} & \multicolumn{1}{c}{\begin{tabular}[c]{@{}c@{}}Dv at \\ 4 sigma\end{tabular}} & \multicolumn{1}{c}{\begin{tabular}[c]{@{}c@{}}SNR at \\ 4 sigma\end{tabular}} \\ \hline
    % \hline\hline
    Planet Model & Grid case & \dv\ Limit  & SNR Limit \\ 
     &  & [\kms]&  \\ \hline \hline
    % \multirow{2}{*}{WASP-33  b} & fixed-phase   & $33.4 \pm 0.1$ & $342 \pm 2$ \\
    %   & fixed-\kp{} & $29.9 \pm 0.1$ & $372 \pm 2$ \\
    % \hline
    % \multirow{2}{*}{HD 189733  b} & fixed-phase   & $46.7 \pm 0.3$ & $663 \pm 4$ \\
    %   & fixed-\kp{} & $47.2 \pm 0.3$ & $808 \pm 5$ \\
    % \hline
    % \multirow{2}{*}{HD 149026  b} & fixed-phase   & $52.9 \pm 0.4$ & $957 \pm 5$ \\
    %   & fixed-\kp{} & $61.6 \pm 0.6$ & $1196 \pm 6$ \\
    % \hline
    % \hline
    
    \multirow{2}{*}{WASP-33  b} & fixed-phase   & $33.4 \pm 0.1$ & $342 \pm 2$ \\
      & fixed-\kp{} & $29.7 \pm 0.1$ & $374 \pm 2$ \\
    \hline
    \multirow{2}{*}{HD 189733  b} & fixed-phase   & $46.7 \pm 0.3$ & $663 \pm 4$ \\
      & fixed-\kp{} & $47.3 \pm 0.4$ & $807 \pm 5$ \\
    \hline
    \multirow{2}{*}{HD 149026  b} & fixed-phase   & $52.9 \pm 0.4$ & $957 \pm 5$ \\
      & fixed-\kp{} & $61.6 \pm 0.6$ & $1196 \pm 7$ \\
    \hline
    \hline
        
    % \multirow{2}{*}{WASP-33  b} & fixed-phase   & $21.2 \pm 0.1$ & $310 \pm 1$ \\
    %   & fixed-\kp{} & $19.7 \pm 0.1$ & $317 \pm 1$ \\
    % \hline
    % \multirow{2}{*}{HD 189733  b} & fixed-phase   & $38.9 \pm 0.2$ & $680 \pm 3$ \\
    %   & fixed-\kp{} & $38.4 \pm 0.2$ & $716 \pm 3$ \\
    % \hline
    % \multirow{2}{*}{HD 149026  b} & fixed-phase   & $46.7 \pm 0.3$ & $987 \pm 4$ \\
    %   & fixed-\kp{} & $45.7 \pm 0.4$ & $1027 \pm 4$ \\
    % \multirow{2}{*}{WASP-33  b} & fixed-phase   & $21.3 \pm 0.1$ & $309 \pm 1$ \\
    %   & fixed-\kp{} & $19.7 \pm 0.1$ & $316 \pm 1$ \\
    % \hline
    % \multirow{2}{*}{HD 189733  b} & fixed-phase   & $38.7 \pm 0.2$ & $670 \pm 3$ \\
    %   & fixed-\kp{} & $38.4 \pm 0.2$ & $726 \pm 3$ \\
    % \hline
    % \multirow{2}{*}{HD 149026  b} & fixed-phase   & $45.5 \pm 0.4$ & $991 \pm 4$ \\
    %   & fixed-\kp{} & $45.8 \pm 0.4$ & $1026 \pm 4$ \\
    % \hline
    % \hline
    % \multirow{2}{*}{WASP-33  b} & \kp-varying   & $17.5 \pm 0.1$ & $242 \pm 1$ \\
    %   & phase-varying & $15.6 \pm 0.1$ & $251 \pm 1$ \\
    % \hline
    % \multirow{2}{*}{HD 189733  b} & \kp-varying   & $28.8 \pm 0.2$ & $497 \pm 3$ \\
    %   & phase-varying & $27.6 \pm 0.2$ & $528 \pm 3$ \\
    % \hline
    % \multirow{2}{*}{HD 149026  b} & \kp-varying   & $30.3 \pm 0.2$ & $724 \pm 3$ \\
    %   & phase-varying & $29.8 \pm 0.3$ & $749 \pm 3$ \\
    % \hline
    % \hline
    \end{tabular}
\end{table}

\section{Discussion}\label{sec:disc}
% \subsection{Planet Model Differences}
% We compare the simulated observations based on three planet models with different atmospheric composition and temperature. 
Our results demonstrate a clear region of significant detection at high \dv\ and SNR for each planet model, which can be clearly seen in Figures \ref{fig:small_grid} and \ref{fig:grids}.
We confirm that a velocity shift \dv\ larger than $3\times$ the instrument resolution ($\sim9$ \kms\ in this case) is necessary to ensure that the planet signal is not removed during the median division (and data detrending) of the total spectrum.
Furthermore, higher SNR provides greater robustness in the detection significance because noisy observations can result in false detections at invalid velocity parameters.
We discuss the accuracy of the detection strength estimation contour fit in Section \ref{sec:disc-limits}.
We compare the two grid cases in Section \ref{sec:disc-gridcases} and compare the differences in planet models in Section \ref{sec:disc-planetmodels}.
In Section \ref{sec:disc-maxsig}, we briefly discuss the alternative method of using the maximum significance value nearby the true velocities as the detection strength.
Finally, we discuss other factors that impact simulated detection estimates in Section \ref{sec:disc-other} and the ease of running the simulations for different planet models and observational parameters in Section \ref{sec:disc-simeff}.

\subsection{Detection Strength Estimated Fit} \label{sec:disc-limits}
By fitting contours as described in Section \ref{subsec:100x100}, we can determine the \dv\ and SNR limits for a given planet model. 
The contour function (Equation \ref{eq:contour_eq}) assumes asymptotic limits at high \dv\ and SNR, which allow us to place lower limits for a specific detection strength.
For a high SNR target, it is expected that the planet signal will completely divide out {below a certain velocity shift during the median division and detrending process}, rendering the planet undetectable.
Similarly, for a planet with a high velocity shift, there should be a minimum SNR below which the planet signal is indistinguishable from noise and cannot be detected. 
In this latter case, the cross-correlation is unable to effectively separate the planet signal at high levels of noise.
{Furthermore, while increasing the number of frames would generally improve detection strengths, because our grids change the total SNR, which is divided over the number of frames, detection strengths are not improved in the fixed-\kp{} grid.}
% For the fixed-\kp{} simulations, the large number of frames at high \dv{} would not improve the detection strength estimate because we simulate over the \textit{total} SNR.
In Figure \ref{fig:grids}, the analytic limits appear to be consistent with the numerical contour limits for all planet models and grid cases, except in the low-\dv{}, high-SNR regime where analytic limits are more conservative. 
Since the simulated region of \dv{} and SNR is a typical region of observational parameters, the contour fit provides reasonably accurate estimates for detection strength limits in \dv{} and SNR.
Hence, we can use these asymptotic limits at high \dv{} and SNR, ensuring that our limits are appropriate for the given region of simulations and typical observations \citep[e.g.][]{finnerty_keckkpic_2023, finnerty_atmospheric_2024}.

However, as mentioned in Section \ref{subsec:100x100}, the contour function deviates from the numerical contours at intermediate \dv{} and SNR values, suggesting a mismatch in the contour fit.
In every grid in Figure \ref{fig:grids}, the contour function fit appears to underestimate the lower detection limit in the intermediate region. 
Hence, the contour function serves only as a rough estimate for detection strengths at these intermediate values, and further investigation of accurate analytic functions is needed for accurate detection limits.

\subsection{Fixed-Phase and Fixed-$K_p$ Grids} \label{sec:disc-gridcases}
\begin{figure}
    \centering
    \includegraphics[width=\columnwidth]{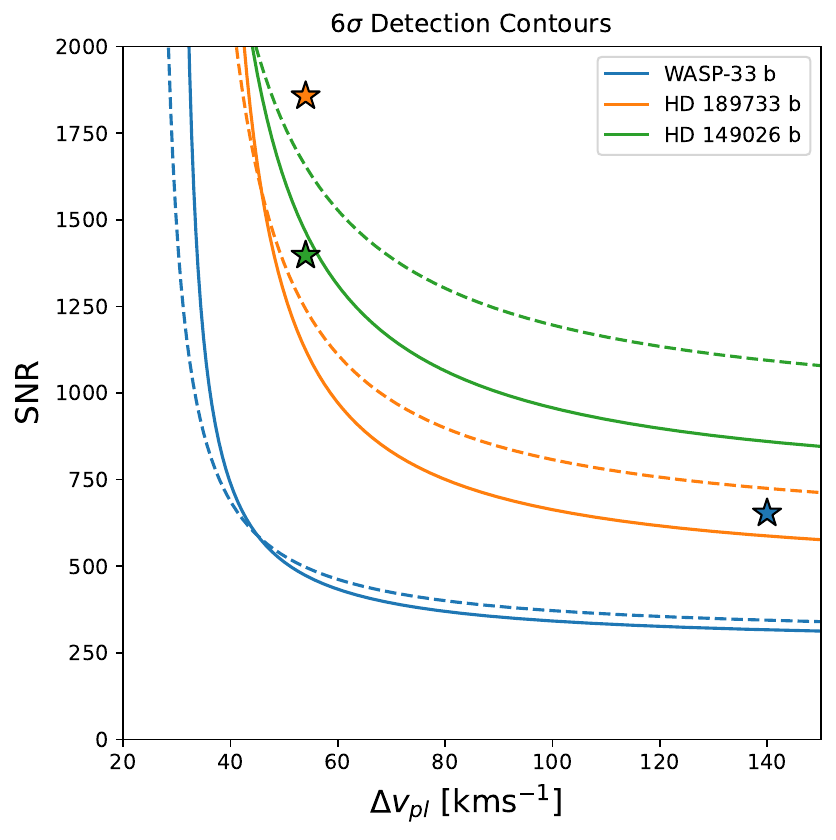}
    \caption{${6}\sigma$ best-fit contour lines from the $100\times100$ detection grids for each planet model and grid case. The fixed-phase grid contours are shown as the solid lines and the fixed-\kp{} grid contours are shown as the dashed lines. The WASP-33 b (ultra-hot Jupiter) model provides lower \dv\ and SNR limits than the other two models. The \dv\ and SNR of Keck/KPIC observations of WASP-33 b, HD 189733 b, and HD 149026 b are also plotted {(as stars)}. The observed parameters {for WASP-33 b and HD 189733 b} fall above the ${6}\sigma$ contours, suggesting that these observations will provide {significant} detections.}
    \label{fig:contours_plot}
\end{figure}

In Figure \ref{fig:grids}, we see that both the fixed-phase grids and the fixed-\kp{} grids give similar contours for all three planet models. 
This can also be seen in Figure \ref{fig:contours_plot} where the ${6}\sigma$ contours for each grid case and planet model are overlaid. 
However, the fixed-\kp{} grids produce slightly higher SNR limits and slightly lower \dv{} limits for all planet models, as shown in Figure \ref{fig:contours_plot} and Table \ref{tab:mins}.
This suggests that the detection strength depends independently on the planet radial velocity semi-amplitude \kp\ and the observed phase range \dphi.
Because the fixed-\kp{} grid varies the total velocity shift by adjusting the total length of the phase range, it may lead to slightly different detection limits.
In the fixed-\kp{} grids, low velocity shifts produce shorter observed phase ranges, but the high \kp\ value of 150 \kms\ produces a sufficient velocity shift that slightly pushes the \dv\ limit down. 
At larger \dv{} and longer phase ranges, the velocity shift deviates further from a linear trend (Eq. \ref{eq:vpl}), which may cause weaker detections compared to the fixed-phase grids and {pushes the SNR limit slightly higher}. 
When planning observations for transiting planets, the planet velocity semi-amplitude \kp{} is known, so it may be more useful to use the fixed-\kp{} simulation grid that varies the velocity shift by adjusting the phase range.
{However, the fixed-phase grids can be used for observations in which the velocity semi-amplitude is not exactly known (e.g. for inclined planet orbits) and for determining a minimum velocity needed for a detection given an observation duration.}
% Overall, the \kp\ parameter appears to have a slightly larger impact in determining the lower \dv\ and SNR limits of detections. 
{Further analysis using different fixed phases and velocity semi-amplitudes is needed to study the impact of each technique in generating detection grids.}
% on the difference between the fixed-phase and fixed-\kp{} grids.

\subsection{Planet Model Comparisons} \label{sec:disc-planetmodels}
The use of three planet models allows us to explore the effect of different atmospheric compositions and temperatures on the detection strength of a planet.
Figure \ref{tab:mins} lists the lower limits for a ${6}\sigma$ detection for each planet model and grid case.

The WASP-33 b (ultra-hot Jupiter) model has the strongest detections and lowest \dv\ and SNR limits among the three planet models, with a minimum $\Delta v_\text{pl} = {29.7} \pm 0.1$ \kms\ and minimum $\textrm{SNR} =  {374} \pm {2}$ for a ${6}\sigma$ detection in the {fixed-\kp{}} grid.
We can clearly see the lower detection limits of the WASP-33 b model in Figures \ref{fig:grids} and \ref{fig:contours_plot}.
This is consistent with the high equilibrium temperature and flux of WASP-33 b relative to its host star.
This allows for better detection of planet lines compared with other planet models, resulting in a high detection even at a low velocity and SNR. 
In particular, the WASP-33 b planet model has the highest planet-to-star flux ratio, which can be seen in Figure \ref{fig:spec}.
Therefore, even in scenarios where observations are short or noise-dominated, an ultra-hot Jupiter can often be detected, due to the large planet-to-star contrast.
In Figure \ref{fig:grids}, the WASP-33 b grids also have the steepest gradient in contour lines such that we see a faster increase in detection strength with increasing \dv{} or SNR compared to the other two models. 
This also attests to the large spectral contrast of the WASP-33 b model whereby small increases in \dv{} or SNR in an observation can significantly improve detection strengths. 

The other two planet models have significantly lower planet-to-star flux ratios compared to the WASP-33 b model.
The HD 189733 b (classical hot Jupiter) model had ${6}\sigma$ detection limits of $\Delta v_\text{pl} = {47.3} \pm {0.4}$ \kms\ and $\textrm{SNR} = {807} \pm {5}$ in the {fixed-\kp{}} grid. 
The HD 149026 b (metal-rich hot Saturn) model had limits of $\Delta v_\text{pl} = {61.6} \pm {0.6}$ \kms\ and $\textrm{SNR} = {1196} \pm {7}$  in the {fixed-\kp{}} grid.
The HD 149026 b model produces a higher SNR limit than the HD 189733 b model, as shown in Figures \ref{fig:grids} and \ref{fig:contours_plot}.
For a sufficient velocity shift, the HD 149026 b model requires a higher SNR to be detected, likely due to its high metallicity which results in weaker planet lines relative to the continuum.
In Figure \ref{fig:spec}, the spectral line depths relative to the star are smaller for the HD 149026 b model, even though the continuum level is higher than the HD 189733 b model.
This results in lower detection strength estimates at low SNR where high noise can obscure the weak lines in the HD 149026 b model.
In the high-SNR regime, the HD 149026 b and HD 189733 b models produce consistent \dv{} limits due to the relatively close flux ratio for the two models.

\subsection{Maximum Significance Estimate} \label{sec:disc-maxsig}
\begin{figure}[ht]
    \centering
    \includegraphics[width=\columnwidth]{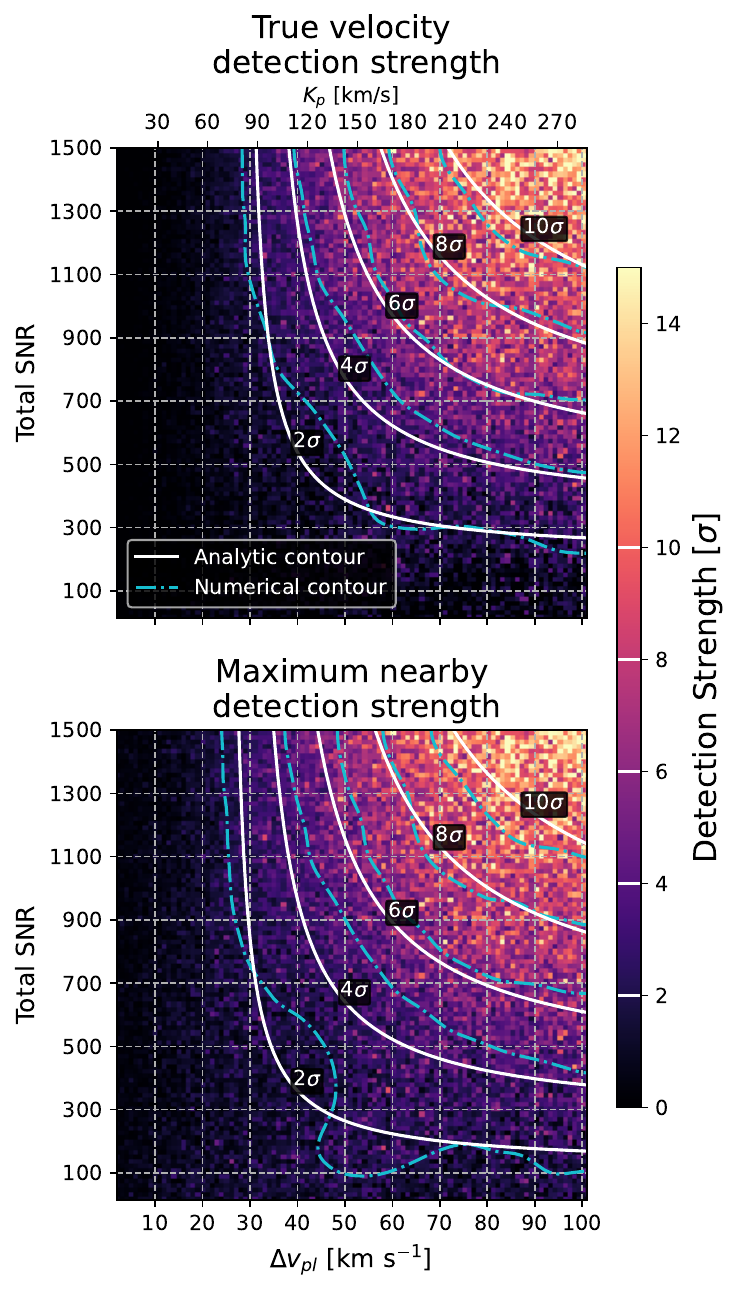}
    \caption{$100\times100$ detection grids for simulated HD 189733 b data, comparing the two methods of determining detection strengths. The upper grid uses the significance at the true velocities of the simulation. The lower grid uses the maximum significance in a simulated \kpvsys\ diagram, given velocity bounds for planet velocity shift in a frame. The detection strength contours are consistent between both methods in the high-\dv{} and high-SNR regime, where the maximum significance is near to the true velocity values. However, using the maximum significance produces a lower detection limit at the low-\dv{} and low-SNR regions. }
    \label{fig:truevmax}
\end{figure}

In Section \ref{sec:ds}, we previously discussed an alternative method of determining detection strengths by taking the maximum significance within a certain range of the true velocities, rather than using the detection strength at the injected planet velocity. 
We show the difference between grids using the true velocity detection strength and the maximum nearby detection strength in Figure \ref{fig:truevmax} for the fixed-phase HD 189733 b model.
For our simulations which assume a perfect model match, the two methods generally produce the same detection strength estimates for sufficiently high \dv{} and SNR, where the peak significance value is found very close to the true velocities.
We can see this in the upper right \kpvsys{} diagrams (\dv{} $\geq50$ \kms, SNR $\geq750$) in Figure \ref{fig:small_grid}, where the maximum value is marginally higher than the value at the true velocities. 
In Figure \ref{fig:truevmax}, both detection strength methods produce similar values above ${\sim}4\sigma$.
However, in the case of a model mismatch, it is possible for the peak to be shifted significantly away from the true velocities.
Then, the true velocity detection strength would not accurately capture the detection significance of a shifted peak in the \kpvsys{} diagram.
For the maximum nearby detection strength, the bounds can be expanded to include a large region of \kpvsys{} space in order to capture the detection peak that is shifted from the true velocities.

In the lower regime of SNR, the maximum nearby detection strength produces significantly higher values than the true velocity detection strength, which can be seen in Figures \ref{fig:small_grid} and \ref{fig:truevmax}. 
The maximum nearby detection strength consistently takes high significance values that represent false detections close to the peak, which are often the result of noise. 
While the true velocity detection strength can still take high values in this region due to noise, it usually captures lower values that raises the detection limits.
Furthermore, in Figure \ref{fig:truevmax}, the analytic contour and numerical contour appear to deviate more when taking the maximum nearby detection strength, suggesting greater mismatch between the analytic contour model to the data.
Hence, in this noise-dominated regime of low \dv{} and low SNR, the maximum nearby detection strength overestimates the detection significance and artificially lowers the detection limit, resulting in inaccurate detections. 

{For simulations with low \dv{}, the artifact at $K_p=0$ dominates over the peak and after the column-wise division of average likelihood values, these detections are suppressed.}
{Hence, both techniques produce similarly low detection strength values in this low-\dv{} regime.}
{At slightly larger \dv{}, the maximum nearby detection strength can still overestimate the detection significance due to false detections in background noise.}

To determine the effects of planet model mismatch on detection strength estimates, the simulations can be modified to include artificial model mismatch between the simulated planet spectrum and the template spectrum.
Then, when determining the detection strength, it may be useful to incorporate both methods such that the true velocity detection strength is used at low \dv{} and SNR while the maximum nearby detection strength is used at high \dv{} and SNR. 
{Doing so will allow for more accurate detections in both the high SNR/\dv{} regime, where model mismatch shifts the peak from the true velocities, and the low SNR/\dv{} regime, where false detections dominate.}
This will likely produce more accurate results than using only one method for the reasons mentioned above.

\subsection{Other Considerations} \label{sec:disc-other}
By using a {typical} ${6}\sigma$ detection limit cutoff, these simulated grids serve as a lower-bound approximation to the detection limits for different planet types. 
However, there were many factors that were ignored in this analysis. 
In our simulations, we assumed an identical and fixed stellar spectrum, planet radius, stellar radius, and orbital period for all planet models and zero systemic and barycentric velocities. 
These assumptions were made to {reduce} the number of independent parameters that may impact the cross-correlation and detection of a given planet.

In particular, the stellar spectrum and planet-to-star area ratio are important factors that impact the overall contrast and thus detection strength, as exemplified by the WASP-33 b model in Figure \ref{fig:wasp_kpvsys}. 
A higher planet-to-star spectral contrast is expected to significantly improve detection strength estimates.
Furthermore, the orbital period and telescope integration time affect the number of frames in a given phase range, which should increase detection strength estimates with more frames.
The systemic velocity and barycentric velocity also affect the total velocity shift of the planet signal, which {may slightly shift} the detection limit on \dv{}.
These parameters should be further investigated with simulations to assess their impact on detection strength estimates.

\begin{figure}[ht]
    \centering
    \includegraphics[width=\columnwidth]{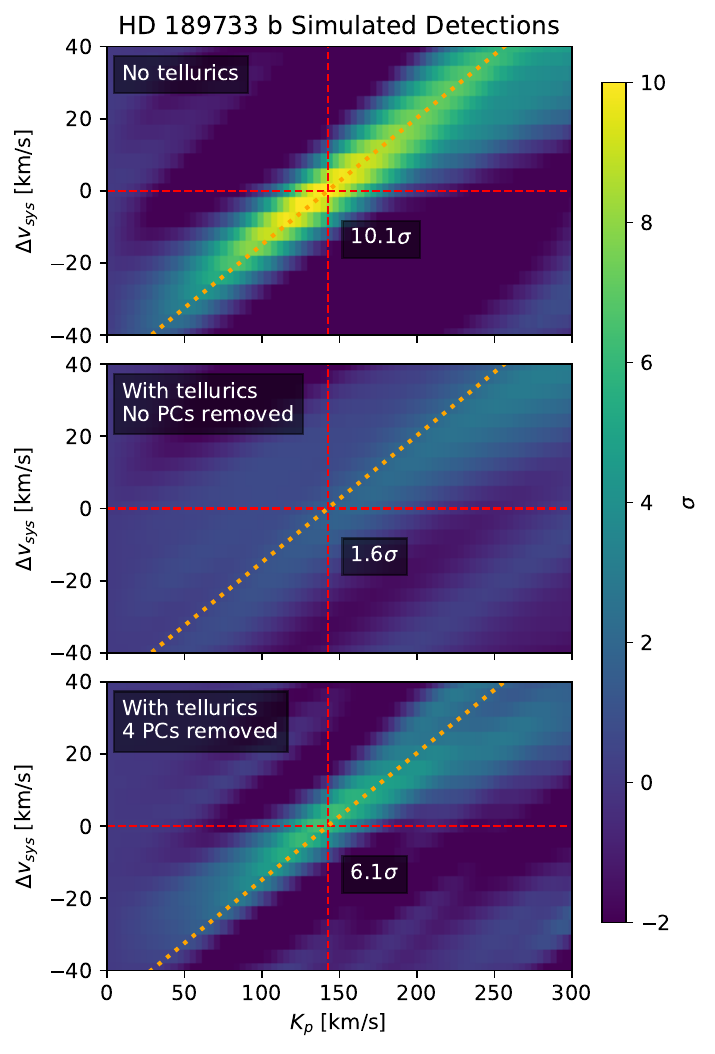}
    \caption{{\kpvsys{} diagrams for simulated HD 189733 b detections with no added time-varying telluric spectrum (top), with added time-varying telluric spectrum but no PCA removal (middle), and with added time-varying telluric spectrum and 4 PCs removed (bottom). Each simulation used $\Delta v_\text{pl} = 50$ \kms{} and SNR$=1000$ with the fiducial phase range for the fixed-phase grid. The simulation with no time-varying telluric features presents the highest detection strength as expected. By adding the time-varying tellurics, the planet becomes undetected until PCA is performed to remove 4 PCs, presenting a more realistic detection strength of $6.1\sigma$.}}
    \label{fig:tellurics}
\end{figure}

{Time-variations in the airmass-dependent telluric spectrum and in the fringing of the KPIC spectrometer \citep{finnerty_atmospheric_2024, finnerty2025a} can significantly impact the detection of exoplanets and require removal by detrending the data using techniques such as PCA/SVD \citep[e.g.][]{cheverall_feasibility_2024}.}
PCA distorts the planet signal, which is typically corrected through a re-injection process \cite[e.g.][]{brogi_retrieving_2019, gibson_detection_2020, line2021, finnerty_keckkpic_2023}. 
{In our simulations, we model a simple time-varying telluric spectrum that is applied to each simulation.}
{We detrend the simulated data via PCA to remove these time-varying effects and study the impact of detrending processes on detection strengths.}
{We show the effect of adding a time-varying telluric spectrum and detrending with PCA on the detection strength in Figure \ref{fig:tellurics}.}
{We find that the addition of time-varying telluric features completely suppresses the planet detection and that performing PCA to remove 4 PCs recovers the planet signal at a lower detection strength (by ${\sim}4\sigma$).}
{With this semi-realistic treatment of time variations, we improve the accuracy of the estimated detection strengths as shown in Figures \ref{fig:wasp_kpvsys} and \ref{fig:contours_plot}.}
{The modeling of time-varying telluric or fringing effects can be extended to include more complicated time-dependence for more accurate detection strength estimates.}
% {By including time-varying tellurics, }
% We also do not consider the impact of telluric effects and fringing effects from the spectrometer, which require removal by detrending the data using techniques such as PCA/SVD \citep[e.g.][]{cheverall_feasibility_2024}.
% However, we do not include the PCA distortion or correction so there may be errors in this replication of the PCA distortion that reduce detection strength.
% PCA is also expected to distort the planet signal, assuming that the PCA stretch is not accurately replicated during data detrending, thus lowering the detection strength of real observed data. 

Despite the many simplifying assumptions in our simulations, we show in Figure \ref{fig:contours_plot} that recent Keck/KPIC observations of the planets {are comparable to the 6$\sigma$ detection limit.} %fall above the conservative ${6}\sigma$ detection limit.
The observed parameter values for WASP-33 b ($\Delta v_\text{pl} = 140$ \kms{}, SNR$=650$) and HD 189733 b ($\Delta v_\text{pl} = 54$, SNR$=1860$) fall in the high-\dv{} and high-SNR regime, respectively, so the analytic contours should serve as accurate detection limit estimates, as discussed in Section \ref{sec:disc-limits}.
Real Keck/KPIC observations of WASP-33 b and HD 189733 b presented ${\sim}12\sigma$ and ${\sim}7\sigma$ detections, respectively, using our detection strength estimation method \citep{finnerty_keckkpic_2023, finnerty_atmospheric_2024}.
{Our analytic contour fit produce respective detection strength estimates of ${\sim}12.1$ and ${\sim}7.9$ at the observed \dv{} and SNR, which are comparable to the real observed detection strength estimate.}
% This overestimation of detection strength is expected given our simplifying simulation assumptions. 
By re-running simulations using more accurate observational parameters (e.g. in Figure \ref{fig:wasp_kpvsys}), we can test the extent to which our simulations can predict true observations.

{However,} the HD 149026 b observed parameters ($\Delta v_\text{pl} = 54$ \kms{}, SNR$=1400$) are slightly below the analytic ${6}\sigma$ threshold, but close to the intermediate values of \dv{} and SNR where the analytic contours were shown to be {slightly} inaccurate (see Section \ref{sec:disc-limits}).
{In other words, the detection grid for the HD 149026 b simulations suggests that the real planet would not be detectable with a $6\sigma$ significance.}
Regardless, observed values of \dv{} and SNR that fall close to the ${6}\sigma$ limit could result in a detection, especially given our simplified simulation assumptions, and using more accurate parameters for simulations may improve results.
For the three planet models, we see that the simulations can serve as an approximate lower limit to a real detection that includes the complex effects mentioned above. 

\subsection{Simulation Efficiency} \label{sec:disc-simeff}
% {Our method of simulating $100\times100$ detection grids enables a reasonable runtime without sacrificing accuracy in simulations.}
{For each simulation, we generate a \kpvsys{} diagram over a relatively small range of \kp{} and \vsys{} with a relatively large step size.}
{Although this may lower the accuracy of the noise estimation and result in slightly less accurate detection strength estimates compared to a larger higher-resolution grid, the speed-up enables a reasonable runtime for the $100\times100$ detection grids.}
{Another method for faster simulations involves cross-correlating each simulated spectrum over a 1D array of total planet velocity shifts for each frame and interpolating the likelihood values onto a \kpvsys{} diagram.}
{However, this method is incompatible with our simulated detrending process and therefore produce very optimistic detection strengths and limits.}
% One of the advantages to this method of simulations is the fast runtime, enabling much larger grids than other approaches.

{With our given method,} each simulation, including generating the data and running the cross-correlation to produce the \kpvsys\ diagrams, runs in ${\sim}${5} minutes on a single CPU slot with 4GB memory, which varies slightly depending on the number of simulated frames.
However, this process is easily parallelizable. 
A $100\times100$ grid (10,000 total simulations) takes ${\sim}${2.5 hours} to run in a cluster environment using an array job that simultaneously performs several simulations as separate tasks on a single compute node with 4GB of memory.
{By improving parallelization using e.g. MPI, the runtime may be reduced further.}
Therefore, these simulations can be generalized to other planet models, stellar models, and system parameters {with fairly little compute time}. 
This is particularly useful for observation planning and feasibility, where the parameters and possible models of a target planet can be used to simulate detections and estimate detection strength estimates for certain parameters (e.g. \dv{} and SNR). 
The simulations can also be used to meaningfully explore non-detection in cross-correlation analyses by estimating expected detection strength based on observational parameters. 
Although the simulations in this work are somewhat optimistic, we can use this framework to study the effects of other parameters on detection strength and make relative comparisons between different planet or stellar models.
% We can use this framework study the effects of other parameters on detection strengths and use the simulations as a guide for observation planning and feasibility. 
% It also allows meaningful exploration of non-detections
Further research should be also conducted {to study more complicated time-varying features} and other likelihood models in the analysis of detection strength estimation. 

\section{Summary and Conclusions}\label{sec:conc}
We simulated high-resolution exoplanet spectra to determine the minimum observational parameters (\dv{} and SNR) necessary for a significant detection (${6}\sigma$). 
We tested three planet models that varied in temperature and composition, based on observations of an ultra-hot Jupiter (WASP-33 b), a classical hot-Jupiter (HD 189733 b), and a metal-rich hot Saturn (HD 149026 b).
We generated simulated spectra for each planet model that varied over the planet velocity shift \dv\ and total signal-to-noise ratio (SNR) in a $100\times100$ grid. 
{These simulations included simple time-varying telluric effects and detrending via PCA to isolate the planet signal.}
For each simulated spectrum, we calculate the log-likelihood over \kp\ and \vsys\ and estimate the detection strength using the significance value at the true velocity parameters of the simulation.
We fit contours to the detection strength grids to estimate the minimum \dv\ and SNR for a ${6}\sigma$ detection. 
In the {fixed-\kp{}} grids, we find ${6}\sigma$ limits {with a 9 \kms\ instrument resolution} of $\Delta v_\text{pl} = {29.7} \pm {0.1}$ \kms\ and $\textrm{SNR} = {374} \pm {2}$ for the WASP-33 b model, $\Delta v_\text{pl} = {47.3} \pm {0.4}$ \kms\ and $\textrm{SNR} = {807} \pm {5}$ for the HD 189733 b model, and $\Delta v_\text{pl} = {61.6} \pm {0.6}$ \kms\ and $\textrm{SNR} = {1196} \pm {7}$ for the HD 149026 b model.
Observed \dv{} and SNR values for existing KPIC observations of each planet were found to be above {or near} the ${6}\sigma$ contours, suggesting that {our detection estimates are fairly accurate}.

The results of this paper provide an optimistic estimate of the lower limits of the planet velocity shift and signal-to-noise ratio required for a significant exoplanet detection. 
These limits can be used for effective observation planning and feasibility of exoplanet targets and serve as a reliability test for future exoplanet detections with high-resolution spectroscopy.
Due to the relatively {fast} speed of simulations, we can expand the simulations to include other planet models and other observational parameters to determine other parameter limits.
However, further research should be conducted on more complex simulations to more accurately determine the lower limits of exoplanet detectability.

\section*{Acknowledgements}
This work was supported by the Anna and Kenneth Taylor Endowment and the Litton Industries Endowment, under the Undergraduate Research Fellows Program at UCLA. 
% This work was supported by the Rex \& Ruth Van Trees Endowment, under the Undergraduate Research Scholars Program at UCLA. 

This work used computational and storage services associated with the Hoffman2 Shared Cluster provided by UCLA Institute for Digital Research and Education’s Research Technology Group. The contributed Hoffman2 computing node used for this work was supported by the Heising-Simons Foundation grant \#2020-1821.

L. F. is a member of UAW local 4811. L.F. acknowledges the support of the W.M. Keck Foundation, which also supports development of the KPIC facility data reduction pipeline. 

% Funding for KPIC has been provided by the California Institute of Technology, the Jet Propulsion Laboratory, the Heising-Simons Foundation (grants \#2015-129, \#2017-318 and \#2019-1312), the Simons Foundation (through the Caltech Center for Comparative Planetary Evolution), and NSF under grant AST-1611623.

This research has made use of the NASA Exoplanet Archive, which is operated by the California Institute of Technology, under contract with the National Aeronautics and Space Administration under the Exoplanet Exploration Program. 

% The data presented herein were obtained at the W. M. Keck Observatory, which is operated as a scientific partnership among the California Institute of Technology, the University of California and the National Aeronautics and Space Administration. The Observatory was made possible by the generous financial support of the W. M. Keck Foundation. The authors wish to recognize and acknowledge the very significant cultural role and reverence that the summit of Maunakea has always had within the indigenous Hawaiian community.  We are most fortunate to have the opportunity to conduct observations from this mountain. 
    
% \end{acknowledgments}


\vspace{5mm}
\facilities{Keck:II(NIRSPEC/KPIC)}


\software{%astropy \citep{astropy:2013, astropy:2018},  
          %\dynesty\ \citep{speagle2020},
          %\texttt{corner} \citep{corner},
          \texttt{scipy} \citep{scipy}
          \petit\ \citep{prt:2019, prt:2020}}



\clearpage
\bibliography{exoplanetbib}{}
\bibliographystyle{aasjournal}



\end{CJK*}
\end{document}